\documentstyle[12pt]{article}
\begin{document}
 
\title {\bf A Binary Approach to the Lorenz Model}
\author{R. L\'{o}pez-Ruiz  \\
                                   \\
{\small  Laboratoire de Physique Statistique} \\
{\small Ecole Normale Superieure} \\
{\small 24 rue Lhomond, 75231 - Paris Cedex 05 (France)}   
\date{ }}

\maketitle
\baselineskip 8mm
 
\begin{center} {\bf Abstract} \end{center}
The order of orbit generation in one-dimensional Lorenz-like maps
is presented within a two letter symbolics scheme. This
order is derived from the natural order of a set of  
fractions associated to the binary sequences. Its relation to the
universal sequence of unimodal maps is explained. 

\newpage
 
\section{Introduction}

One-dimensional maps are used for modelling three-dimensional flows
with large dissipation. The work of labelling and 
ordering periodic orbits in those systems can be 
accomplished by the methods of symbolic dynamics \cite{bailin}.
Thus, for instance, the different periods  embedded 
in a chaotic region can be generated in a systematic 
way with two or more symbols. In the case of unimodal
maps, it is a well-established
 result the ordering of all sequences
($U$-sequence) with a two letter symbolics \cite{metropolis,derrida,collet}.
This unidimensional result has been generalized to an orbit
implication diagram  for three dimensional 
flows and two dimensional orientation preserving maps wich evolve
into horseshoes under parameter variation. This diagram
is independent of dissipation, from the conservative limit (zero dissipation)
to the unimodal limit (infinite dissipation) \cite{lopez}.

In the case of the Lorenz-like flow, a cubic map has been
presented as a simplified model that keeps the antisymmetry
of the Lorenz equations. Its periodic structure
has been investigated with a three letter symbolics \cite{ding}.
Also, families of maps as those shown in reference \cite{sparrow}
(appendix J) model some aspect of the behavior of Lorenz system.
\par

Here we undertake another approach to the problem of orbit
generation in the Lorenz model.
 The following one-dimensional map, 
$x_{n+1}=f_{\lambda}(x_n)$, on the interval $[-1,1]$ is proposed
as a simple model keeping the symmetry properties 
( $f_{\lambda}(x_n)=-f_{\lambda}(-x_n)$) and modelling Lorenz system:
\begin{eqnarray}
 x_{n+1} & = & \lambda x_n+(\lambda -1) \;\;\;\; if \;\;\; -1<x_n<0 \nonumber \\
& & \lambda x_n-(\lambda -1) \;\;\;\; if \;\;\;\;\;\; 0<x_n<1 
\end{eqnarray}
It has been used to determine the order in which the orbits are created
when the parameter $\lambda$ varies on $[1,2]$.
Its dynamics can be understood with
a two letter symbolics (binary sequences of zeroes and ones:
$0$ if $-1<x_n<0$ and $1$ if $ 0<x_n<1$). For $\lambda=1$,
$x_{n+1}=f_{\lambda=1}(x_n)=x_n$, each point on $[-1,1]$
is a fixed point and there is no periodic orbit
with period greater than one.
For $\lambda=2$, $x_{n+1}=f_{\lambda=2}(x_n)$ is the Bernouilli
shift and all periodic orbits are present. Thus each orbit
(binary sequence) is created for a value of the parameter
$\lambda$ on $[1,2]$. This value can be calculated exactly
for the model (1) as a root of a polynome in $\lambda$
of order equal to the period of the orbit. Finally, the sequence
of creation of all periodic orbits is obtained.
That means that an order in the space of binary sequences
is built. This order ($L$-order) is established when a set of fractions
associated to the binary sequences is ordered in a decreasing
natural way.\par

In the case of unimodal maps a similar study has been performed
with the following map $x_{n+1}=g_{\lambda}(x_n)$ on $[-1,1]$
($f_{\lambda}(x_n)=f_{\lambda}(-x_n)$):
\begin{eqnarray}
 x_{n+1} & = & \;\;\lambda x_n+(\lambda -1) \;\;\;\; if \;\;\; -1<x_n<0 \nonumber \\
& & -\lambda x_n+(\lambda -1) \;\;\;\; if \;\;\;\;\;\; 0<x_n<1 
\end{eqnarray}
Each point in the interval $[-1,0]$ is fixed for $\lambda=1$
and there is no other periodic orbit. For $\lambda=2$,
$x_{n+1}=g_{\lambda=2}(x_n)$ is the tent map and
all periodic orbits are present. As done for system (1),
each orbit (binary sequence) is created for a value of the parameter
$\lambda$ on $[1,2]$ that can be calculated exactly
as a root of a polynome in $\lambda$ of order equal 
to the period of the orbit.

The orbits created in system (2) for each value of $\lambda$ are related
to those created for the same value of $\lambda$
in system (1).
A close connection is established between the ordered structure
of periodic orbits of unimodal maps  and that of 
the Lorenz map.
A transformation from the $L$-order to the actual order
($R$-order or $U$-sequence), and viceversa, is presented. \par

In summary, the two orders ($L$-order and $R$-order) introduced
in the space of binary sequences result independent of the
original systems (1) and (2). These orders are consistent,
well-defined and related structures in that space.
In the next sections this point of view is presented:
the binary nomenclature is presented in section 2 and
the $L$-order and $R$-order are explained in sections 3 and 4,
respectively.

\section{Space of Binary Sequences: Nomenclature}
Let us call {\it $\cal B_S$} the space of binary sequences. An orbit of period
$n$ ($O_n$) is formed by a binary sequence non periodic (irreducible) 
of $n$ digits and all sequences built from the iteration of this 
irreducible sequence as a block. It is represented by the irreducible 
orbit. An example can be: 
$O_3=100\equiv \{100,100100,100100100,\ldots \}$. The {\it real equivalent}
of the orbit $O_n=\alpha_{n-1}\alpha_{n-2}\ldots\alpha_1\alpha_0$ is 
the real number $\sum_{i=0}^{n-1}\alpha_i2^i$. The set formed by
an $n$-periodic orbit, $O_n$, and its $n-1$ cyclic permutations will be called
{\it orbital} of period $n$ ($[O_n]$). For example, 
$[O_3]=[100]\equiv \{100,010,001\}$. The set of the conjugate orbits is the
{\it conjugate orbital}: $[\overline{O}_3]=[011]\equiv \{011,101,110\}$.
The number of orbitals $N_n$  of period $n$ is given by:
\begin{equation} 
N_n=\frac{2^n-\sum_{i=1}^{k}m_iN_{m_i}}{n}
\end{equation}
where $\{m_1,m_2,\ldots,m_k\}$ are the integer divisors of $n$ 
excluding $\{n\}$. \par

\section{$L$-Order ("Lorenz" Order)}
If $[O_n]\neq [\overline{O}_n]$, the set $L_n^D=[O_n;\overline{O}_n]\equiv
[O_n]\cup [\overline{O}_n]$ is called {\it $L$-doublet}.
If $[O_n]=[\overline{O}_n]$, the set $L_n^S=[O_n;]\equiv [O_n]$ 
is called {\it $L$-singlet}. Given a $L$-doublet 
or $L$-singlet, the orbit that starts in the left side by $1$  followed
by the subsequence with the smallest real equivalent is called the
{\it characteristic orbit} ($O_n^c$). This orbit is chosen as representantive:
$L_n^D=[O_n^c,\overline{O}_n^c]$, $L_n^S=[O_n^c;]$. For example,
if $O_3^c=100$ then $L_3^D=[100;011]$ or if $O_4^c=1001$ then $L_4^S=[1001;]$.
The {\it associated fraction} ($r$) of a $L$-doublet or a $L$-singlet 
is a fraction associated to its caracteristic orbit,
$O_n^c=\alpha_1\alpha_2\ldots\alpha_n$ and defined as:
\begin{equation} 
r=\left(\sum_{i=1}^{n}\frac{\alpha_i}{2^i}\right)\cdot\frac{2^n}{2^n-1}
\end{equation}
These fractions are (except the trivial $1$-periodic 
orbit with $r_{[1;0]}=1$) in the range 
$r_{[10;]}=\frac{2}{3}>r>\frac{1}{2}=r_{[100\stackrel{\infty}{\ldots};
011\stackrel{\infty}{\ldots}]}$. \newline
Now we establish an order relation
in the space of $L$-doublets and $L$-singlets ($[;]$),
called {\it implication}
and represented by the symbol $\Rightarrow$.
We will said that the orbit set $[;]_i$ implies $[;]_j$ if 
the associated fractions verify $r_i<r_j$:
\begin{displaymath}
[;]_i \Rightarrow [;]_j \leftrightarrow r_i<r_j
\end{displaymath}
For example, $L_4^D\equiv [1000;0111]\Rightarrow L_3^D\equiv [100;011]$
because $r_{[1000;0111]}=\frac{8}{15}<\frac{4}{7}=r_{[100;011]}$.
Following these rules the ordered binary set showed in table 1
has been found, where an orbit set implies all the orbit set above it (Fig. 1). 
We represent the space of binary sequences and the $L$-order relation
by $({\cal B_S}, {\cal L_{\Rightarrow}})$. \par

\section{ $R$-Order("R\"ossler" Order; $U$-sequence)}
Another order type, we call R-order, is established in $\cal B_S$.
This order generates the $U$-sequence. It is built from
the $L$-order with the following steps.
We group in a new way in $({\cal B_S},{\cal L_{\Rightarrow}})$: 
each $L$-doublet, $L_n^D=[O_n^c;\overline{O}_n^c]$, and 
its doubled $L$-singlet, $L_{2n}^S=[O_n^c\overline{O}_n^c;]$, 
is grouped in a {\it $L$-triplet}, 
$L_n^T=[O_n^c,\overline{O}_n^c;O_n^c\overline{O}_n^c]=L_n^D\cup L_{2n}^S$.
$L$-singlet does not belong to any $L$-triplet is called $L$-singlet$_d$.
They have the property: $L_{2n}^{S_d}=[O_n\overline{O}_n;]\rightarrow 
O_n=O_{\frac{n}{2}}\overline{O}_{\frac{n}{2}}$. Then, $\cal B_S$ is 
divided in a new partition: $L$-triplets and $L$-singlets$_d$. 
We define the binary sequence transformation $\cal F_{B_S}$ 
that transforms the binary sequence
$\{l_1l_2\ldots l_n\}$ in the binary sequence 
$\{r_1r_2\ldots r_n\}$ according to the law:
\begin{eqnarray}
r_{i+1} & = & l_i+l_{i+1}\pmod{2} \\
l_0     & = & l_n  \nonumber
\end{eqnarray}
The inverse transformation ${\cal F_{B_S}}^{-1}$ is defined as follows:
\begin{equation}
l_i=\sum_{j=i}^{i}r_j
\end{equation}
If we apply $\cal F_{B_S}$ to the characteristic orbit $O_n^c$ of a $L$-triplet
the result is the regular-orbit $O_n^r$, an orbit with even number of $1$.
If $\cal F_{B_S}$ is applied to the orbit $O_n^c\overline{O}_n^c$ 
of the same $L$-triplet the result is the flip-orbit $O_n^f$, 
an orbit with odd number of $1$. Thus the $L$-triplet is transformed by
$\cal F_{B_S}$ in two independent orbitals whose union, 
$R_n^D=[O_n^r;O_n^f]\equiv
[O_n^r]\cup [O_n^f]$, is called {\it $R$-doublet}.
The transformation of the $L$-singlet$_d$ $L_{2n}^{S_d}$ by $\cal F_{B_S}$ 
produce an orbital said {\it $R$-singlet}, $R_n^S$. The action of
$F_{S_B}$ is summarized as follows:
\begin{eqnarray*}
{\bf \cal F_{B_S}}:\;\;\;\; {\cal B_S} & \longrightarrow & {\cal B_S} \\
L_n^T\left\{\begin{array}{c}
L_n^D \\
L_n^S
\end{array}\right.          & 
\begin{array}{c}
\longrightarrow \\
\longrightarrow 
\end{array}                 & 
\left.\begin{array}{c}
[O_n^r] \\
\,[O_n^f]
\end{array}\right\} R_n^D   \\
L_{2n}^{S_d} & \longrightarrow & R_n^S
\end{eqnarray*}
Moreover, $\cal F_{B_S}$ transfers the $L$-order in 
($\cal B_S$, $\cal L_{\Rightarrow}$) to a new order said R-order
(Table 2 and Fig. 2).
This order relation is defined as follows: if 
$L_{n_i}^{S_d,T}\Rightarrow L_{n_j}^{S_d,T}$ then the transformed orbitals
by $\cal F_{B_S}$ verify: 
$R_{n_i}^{D,S}\Rightarrow R_{n_j}^{D,S}$.(This is well defined because
the L-doublet and the L-singlet composing the L-triplet do not have
any other orbital between them in the $L$-order).
The new ordered space is denoted by ($\cal B_S$, $\cal R_{\Rightarrow}$).
This is the $U$-sequence (see Table 1). \par.

\section{Conclusion}

We have presented two natural binary structures.
One of them corresponds to the order of orbit generation in Lorenz-like
maps (within a two letter symbolics scheme).
This order is derived from the
natural sequence of a set of fractions associated to those periodic
orbits (binary sequences).  The other one is the well-known universal
sequence in unimodal maps. There is a close relation between 
these two ordered sets. The transformations from one order
to the other are presented. \par

{\bf Acknowledgements:}
I thank Prof. R. Gilmore (Philadelphia) and Prof. G. Mindlin (Buenos Aires)
for useful discussions and the Spanish Gouvernement 
for a postdoctoral research grant.

\newpage
\begin{center} {\bf Figure Captions} \end{center}

{\bf Fig 1.} Number of orbits, $N(r)$, as a function of the associated
fraction, $r$. ($N(r_0)$ represents the number of orbits
with period $n\leq 30$ whose associated fraction $r$ verifies 
$r_0-\Delta r<r<r_0+\Delta r$, with $\Delta r\sim 10^{-5}$).

{\bf Fig 2.} Three dimensional interpretation of a $L$-triplet,
$L_4^T=[1000,0111;10000111]$. This
could be understood as a three-orbit symmetry bifurcation in the Lorenz
attractor. ($\cal F_{B_S}$ could be interpreted as a relation 
between [saddle-node, period doubling] bifurcations in R\"ossler
attractor and [three-orbit symmetry, period doubling] bifurcations
in Lorenz attractor, respectively).

\begin{center} {\bf Table Captions} \end{center}

{\bf Table 1.} $R$-order and $L$-order
({$\cal R_{\Rightarrow}$}, {$\cal L_{\Rightarrow}$}) 
in the set of binary sequences until
period $n=6$, associated fractions ($r$), and order of the orbits 
in unimodal maps.($^*$ means period-doubled orbits).

{\bf Table 2.} The two different orbit groupings, "R\"ossler" and "Lorenz" type, 
established in the space of binary sequences $(\cal B_S)$.

\newpage
\begin{table}[htb]
\begin{center}
\begin{tabular}{|l|c|l|c|c|} \hline
$R-Order$  & $Period_{order}$  & $L-Order$ & $r$ & $r$ \\ \hline\hline 

$0      $   & $ 1_1  $  & $1$ ; $0$          & $1       $& $1.00000$  \\ 
$1      $   &           & $10          $     & $2/3     $& $0.66666$  \\ \hline
$10     $   & $ 2_1^*$  & $1001        $     & $3/5     $& $0.60000$  \\ \hline
$1011   $   & $ 4_1^*$  & $10010110    $     & $10/17   $& $0.58823$  \\ \hline
$101110 $   & $ 6_1  $  &$100101$ ; $011010$ & $37/63   $& $0.58730$  \\ 
$101111 $   &           & $10010101101 $     &$2394/4095$& $0.58461$  \\ \hline
$10111  $   & $ 5_1  $  & $10010$ ; $01101$  & $18/31   $& $0.58064$  \\ 
$10110  $   &           & $1001001101$       & $589/1023$& $0.57575$  \\ \hline
$101    $   & $ 3_1  $  & $100$ ; $011$      & $4/7     $& $0.57142$  \\ 
$100    $   &           & $100011$           & $35/63   $& $0.55555$  \\ \hline
$100101 $   & $ 6_2^*$  & $100011011100$     &$2268/4095$& $0.55384$  \\ \hline
$10010  $   & $ 5_2  $  & $10001$ ; $01110$  & $17/31   $& $0.54838$  \\ 
$10011  $   &           & $1000101110$       & $558/1023$& $0.54545$  \\ \hline
$100111 $   & $ 6_3  $  & $100010$ ; $011101$& $34/63   $& $0.53968$  \\ 
$100110 $   &           & $100010011101$     &$2205/4095$& $0.53846$  \\ \hline
$1001   $   & $ 4_2  $  & $1000$ ; $0111$    & $8/15    $& $0.53333$  \\ 
$1000   $   &           & $10000111$         & $135/63  $& $0.52941$  \\ \hline
$100010 $   & $ 6_4  $  & $100001$ ; $011110$& $11/21   $& $0.52380$  \\ 
$100011 $   &           & $100001011110$     &$2142/4095$& $0.52307$  \\ \hline
$10001  $   & $ 5_3  $  & $10000$ ; $01111$  & $16/31   $& $0.51612$  \\ 
$10000  $   &           & $1000001111$       & $527/1023$& $0.51515$  \\ \hline
$100001 $   & $ 6_5  $  & $100000$ ; $011111$& $32/63   $& $0.50793$  \\ 
$100000 $   &           & $100000011111$     &$2079/4095$& $0.50769$  \\ \hline
\end{tabular}
\caption{}
\end{center}
\end{table}

\begin{table}[htb]
\begin{center}
\begin{tabular}{|c|c|} \hline
$"Lorenz"$ $Grouping$           &  $"R\ddot{o}ssler"$ $Grouping$\\ \hline\hline 

Binary                             &    Binary               \\
Sequences                          &    Sequences            \\ \hline
Orbits, $O_n$                      &    Orbits, $O_n$        \\
Orbitals, $[O_n]$                  &    Orbitals, $[O_n]$    \\ \hline
$L$-doublet, $L_n^D$               &                         \\
$L$-singlet, $L_n^S$               &                         \\ \hline
$L$-triplet, $L_n^T$               &    $R$-doublet, $R_n^D$ \\
$L$-singlet$_d$, $L_{2n}^{S_d}$    &    $R$-singlet, $R_n^S$ \\ \hline
\end{tabular}
\caption{}
\end{center}
\end{table}

\end{document}